# Student's Video Production as Formative Assessment

Eduardo Gama[1], Marta F. Barroso[2]

[1] Colégio Pedro II, Rio de Janeiro, Brazil

[2] Universidade Federal do Rio de Janeiro, Physics Institute, Rio de Janeiro, Brazil

Paper presented at ICPE 2013 – International Conference on Physics Education, Prague. Submitted for publication in the Annals of ICPE.

This work was done under a grant from CAPES – Observatório da Educação / Brazil.

Abstract

*Learning assessments are subject of discussions that envolve theoretical and practical approaches. To measure learning in physics by high school students, either qualitatively or quantitatively, is a process in which it should be possible to identify not only the concepts and contents students failed to achieve but also the reasons of the failure. We propose that students' video production offers a very effective formative assessment to teachers: as a formative assessment, it produces information that allows the understanding of where and when the learning process succeeded or failed, of identifying, as a subject or as a group, the defficiencies or misunderstandings related to the theme under analysis and their interpretation by students, and it provides also a different kind of assessment, related to some other life skills, like the ability to carry a project to its conclusion and to work cooperatively. In this paper, we describe the use of videos produced by high school students as an assessment resource. The students were asked to prepare a short video, which was then presented to the whole group and discussed. The videos reveal aspects of students' difficulties that usually do not appear in formal assessments such as tests and questionnaires. After the use of the videos as an component of classroom assessments and the use of the discussions to rethink learning activities in the group, the videos were analysed and classified in various categories. This analysis showed a strong correlation between the technical quality of the video and the content quality of the students argumentation. Also, it was shown that the students don´t prepare their video based on quick and easy production, they usually choose forms of video production that require careful planning and implementation, and this reflects directly on the overall quality of the video and of the learning process.*

# Introduction

Teaching and learning – to find out if the connection was made it is necessary to assess learning. In most cases, teachers do not think extensively about how the assess learning; they basically do what they have previously experienced.

But assessment can be a fundamental tool in the learning process. If taken as a formative assessment tool [1], it is possible to retrace steps in teaching, to rethink classroom activities and developments, therefore improving the learning capabilities of the students.

In general, the teaching process does not aim uniquely at content subject learning. It envolves skills related to the interaction with peers, with autonomy and intellectual independence, and somehow with the complection of projects and actions, as happens in real life. But assessments in general do not take these complementary and important aspects into account. It is very difficult to develop an action with pencil and paper in a limited period of time.

This paper presents an activity used as a formative assessment tool in high school physics education. The students were asked, after formal instruction, to produce a short video on one of the themes studied during term. They have to work in groups, and also write the conception and production mechanisms of the video.

After the videos were produced, the teacher watches all of them and prepares a video presentation and discussion session. The discussion includes all the students, and the process grades the student with a small part of the final grade.

The posterior analysis of the videos reveal what was an unexpected product of the activity: they produce a fine assessment tool, for they reveal some aspects which cannot be easily assessed in content learning. The videos were analysed in a series of aspects, and the results allow us to conclude that video production by high school students can be used with good results to assess learning.

# Preliminary considerations

The new information and communication technology has profound impact on young students. They deal with music, videos, and communication in new ways that are mostly not present at school activities.

The acessibility of mobile phones, tablets, and video cameras to almost every house, even in not rich regions over the world, poses some new possibilities to physics teaching. One of them is the use video production in classrooms. According to Vonk [2],

> *"Like it or not, we seem to be using video in almost exactly the same way that we have used writing. And like it or not, a video analog has saddled up next to virtually every form of writing known, except in academia, where most professors I know are still requiring only written work."*

The use of videos in classroom has been subject of many discussions. One of the most interesting ones is related to the use as an approach to laboratory experiments [3].

But there is another possibility, related to learning assessments. The definition of learning is related to the complex construct of understanding. To have a measurement of understanding, an operational definition of it is necessary – and any operational definition needs to be broad enough so that the concept of learning and understanding is not restricted to answering a few simple questions on concepts, or solving some standard exercises, or completing some preordered activity. According to White and Gunstone [4],

> *"We contend that assessment in schools is too often narrow in range. The oral questions that teachers ask in class and their informal and formal written tests usually are confined to requiring short answers of a word or two or a number, a choice from a few alternatives, or 'essays' of various lengths. While there is nothing wrong with these tests, they are limited in type. Limited tests provide a limited measure of understanding, and, worse, promote limited understanding. We advocate use of diverse probes of understanding as an effective means of promoting high quality learning."(p. vii)*

In particular, when it comes to formative assessment, that is, the assessment during the course, which is intended to diagnose the comprehension of the themes, the progress the students are making, and with that have some tools to reorganize students' learning and teaching activities, video production is a very interesting tool.

In making a video, students have to access their cognitive resourses e define the strategies that allow them to fulfill what was asked by the teacher. This activity has some meta-cognitive characteristics, for it makes the students think about their actions, planning and replanning them, trying different language forms till they find the appropriate one, recognizing and overcoming their limitations in the process of production. It is an assessment tool for the teacher, and much more than that: it is a learning resource, complementing the ones used by teachers. Also, the aspect of socialization between the students, provided by the audiovisual format, can provide new questions, new difficulties and reveal aspects of concepts and contents that were not clear.

Another characteristic of video production by students is that it can be though as a means to acquiring meaningful learning in the sense of Ausubel's theory [5]. According to it, any new information has to interact with a previous information already present in the cognitive structure of the student. The teacher needs to know what his students know, so that he can provide activities that allow the assimilation and reorganization of the cognitive structure of them. In producing a video on a physics theme, the students have to access their cognitive resourses frequently; the production implies a ressignification of concepts and reorganization of the cognitive structure, specially when the student has to work collaboratively. Also, it is easier for the teacher to evaluate if the learning was a mechanical one. This evaluation occurs in three occasions: in the video production, when students ask the teacher for help, during the analysis of the video by the teacher, and in the classroom discussion of all the videos, when the students have to interact with other groups.

## Description of the assessment process with the students

The use of video production by students as an assessment tool was developed in a high school in Rio de Janeiro. This school, Colégio Pedro II, named after the second and last emperor of the country, is one of the federal and traditional institutions of basic education in Brazil. The mechanism by which students are accepted in this Colégio is chance, and this means that classes have a multicultural and multieconomic profile.

The themes in physics studied presented to students in high school first year (14-15 years) were geometrical optics and heat, in second year (15-16 years) introductory mechanics, and third year (16-17 years) electricity, oscillations and waves, and sound.

The teacher has some concepts to explore every term, discussed in the physics team of teachers. The presentation of the themes is mainly done in traditional ways: classroom activities based on lecturing, exercises, videos and discussions.

After the term (three months), the students were asked to produce a video that should be used as part of their grades on the term. They should gather into groups of no more than 5, choosen by themselves, and prepare a video on one of the subjects studied. There were no constraints other than the duration (between 1 and 10 minutes) and the character of being a presentation to colleagues. The format, the subject, the means were all their choice. And they should also write a short paper, of no more than 2 pages, describing what wasw produced, how it was produced, with a brief explanation of the physics envolved.

The teacher collects and watches all the videos, and reads all the written texts. He or she presents comments and corrections to the written texts, and prepare comments and discussions on the videos. The analyses involve objective aspects like the technical format, the physics envolved, and the connection between the proposal and the final video. Special care was taken the physical content of the theme.

Finally, there was a video session for the classroom. In this opportunity, the teacher uses every chance to improve the learning of the concepts and the interaction of the students with themselves and with physics.

## The analysis of the videos

The videos were gathered and analysed. The aspects choosen for the analysis were related to the research question: is video production by students a reliable assessment tool?

With this in mind, the videos were analysed on three main categories: the physics content presented, the format choosen to present this content, and general aspects.

The content of the video regarded basically if the physics involved was correct, if the presentation was clear, if it was compatible with the proposal. The technical format was divided into the type used (a movie, a superposition of slides, an experiment, a cartoon, and some mixed types. The general aspects are relate to questions like if the video showed internal logic and / or internal coherence and consistency, if there is an activity like an experiment to present some physics phenomena, if there was an explanation of the results of the experiment and how was the quality of the argumentation.

In this paper, it is presented the results obtained with classes during the year of 2011. In Table 1, we present the global data: 55 videos were produced, and 232 students participate (131 female, 101 male), divided by year in school.

Table 1 – The students involved and videos produced in each group

|  | videos | students |
|---|---|---|
| 1st year high school | 12 | 55 |
| 2nd year high school | 22 | 98 |
| 3rd year high school | 19 | 75 |
| missing | 2 | 4 |
| Total | 55 | 232 |

In Table 2, we show the division of the videos on themes. One can notice that elementary dynamics is the theme that most videos treat; and this can be seen by the majority of students in 2nd year students, and their main theme is introductory mechanics.

Table 2 – The themes of the videos on introductory physics

| dynamics | 37 (67%) |
|---|---|
| geometrical optics | 7 |
| thermal physics | 2 |
| mechanical waves | 3 |
| electricity | 1 |
| waves and optics | 2 |
| contemporary issues | 1 |
| generalities | 2 |

In connection to the assessment of the physical content, it was observed that 35% of the productions were correct and 45% were partially correct. Also, the physical ideas were presented clearly in almost half of the videos (47%), the connection between proposal and product was satisfactory in 75% of the videos, and there was a logical sequence in the videos in 87%; these data are shown in Tables 3 and 4.

Table 3. The videos analysed in connection to the physical content.

| The physical content is correct | | | The presentation is clear | | |
|---|---|---|---|---|---|
| correct | 19 | 35% | yes | 26 | 47% |
| partially correct | 25 | 45% | no | 7 | 13% |
| incorrect | 8 | 15% | partially | 21 | 38% |
| not applicable | 3 | 5% | not applicable | 1 | 2% |

Table 4. The videos analysed in connection to the physical content.

| There is a connection between proposal and product | | | There is a logical sequence | | |
|---|---|---|---|---|---|
| yes | 41 | 75% | yes | 48 | 87% |
| no | 3 | 5% | no | 0 | 0% |
| partial | 8 | 15% | partial | 5 | 9% |
| not applicable | 2 | 4% | not applicable | 2 | 4% |

The videos were presented in a series of formats; about 33% of them were movies, 27% were a combination of videos and slide presentations, and the rest was presented as a theater scene, comics, only slide presentation in sequence, etc. Almost all of them (84%) used music, but of these music was just background (93%) and not part of the story.

It can also be noticed that although in only 22% of the videos the students prepared an experimental situation, there was some experiment shown in 64% of them, as shown in Table 5.

Table 5. The use of experiments in the images.

| There was preparation of experiment. | | | An experiment was filmed. | | |
|---|---|---|---|---|---|
| yes | 12 | 22% | yes | 35 | 64% |
| no | 43 | 78% | no | 20 | 36% |

The surprising aspect was the correlation observed between the technical quality of the video and the exactness of the physics discussed: only 7% were technically poor. 42% of the videos were technically very well done; and from these, 90% were entirely conceptually correct.

## What the videos reveal about learning physics

The videos provided a very useful assessment tool. The students were given the possibility of talking physics, and in it they revealed what it takes a long and hard way for the teachers to find ou. In general, this kind of evaluation is only possible with long individual interviews, or carefully prepared (with the right questions) questionnaires.

As an example of a video with reveals difficulty in learning, there is a video, "Sports and Physics". The image of the video is a 100m man race, with U. Bolt winning. A small part of his words are

> "In an atletic race, the athlete shall keep his body upright while he completes the curve. (…) In this case, the centripetal force he produces while on the curve using the incline acts against the centrifugal force that sends him outwards."

In this example, the whole text presents many incorrect conceptions, and it can be noticed the confusion about concepts related with elementary dynamics (on inertial forces, centripetal acceleration, third law). Also, this video is presented as a reproduction of a TV news, technically careless in the production.

Another example shows what can be obtained: in a video named "Law of Gravity", a female student receives his exam graded zero, and a friend teaches her about gravity by rolling down the stairs. She says,

> ''What is gravity? According to Newton's law, not this Newton, Isaac Newton, gravity is the force of attraction that material bodies exert on one another. (...) Loosely speaking, it is the law of physics that hold things attached to Earth. It is what makes this... happen."

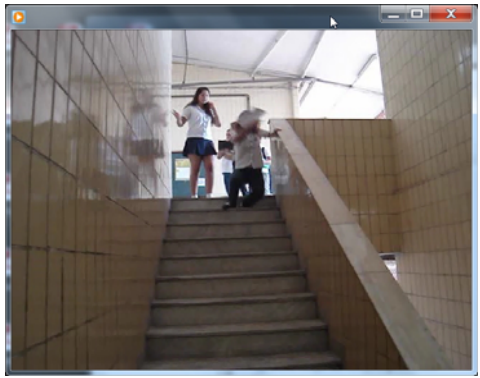

Figure 1. Law of Gravity

In this situation, it can be seen that the students are amusing themselves, and using a language that clearly corresponds to their age. They make a joke with one of the authors, named Newton, and with rolling down the stairs, as shown in Figure 1.

# Conclusions

This paper proposes that the production of short physic videos is a very appropriate formative assessment tool.

The use of mobile phones, tablets and cameras are disseminated nowadays, and students do use them often. In school, the teachers are still reluctant to understand the possibilities of use of these tools as part of their teaching materials, as suggest by Vonk [2].

In fact, videos can be used for data collection in physics and in physics teaching. And also can provide a very useful assessment tool, another kind of probe as proposed by White and Gunstone [4].

The proposed activity is an extra class activity, producing a video and relating the production. This activity requires more skills than just learning physics: requires cooperative team work, active participation of the students in their proces of learning [5], and are related to the use of technology they are familiar with.

The videos allowed the teacher to check precisely how the students interpreted what he or she taught, still in time to promote changes. The videos revealed how the students think about the topics, and surprisingly they seem to have spent time in preparing the videos. It was noticed that there is a strong correlation between the high technical quality and the quality of content.

The main conclusion is that the use of video production by students in physics high school classes is a possible and reliable formative assessment tool.